\documentclass[preprint,12pt]{elsarticle}

\usepackage{graphicx}

\usepackage{amssymb}

\journal{Physics Letters B}
\usepackage{amsmath}
\usepackage{braket, slashed}
\usepackage{ctable}
\usepackage[table]{xcolor}

\definecolor{grey}{rgb}{0.5,0.5,0.5}
\definecolor{lgrey}{rgb}{0.9,0.9,0.9}
\definecolor{LightRed}{rgb}{1.0,0.5,0.5}

\newcolumntype{L}[1]{>{\raggedright\hspace{0pt}}p{#1}}
\newcolumntype{C}[1]{>{\centering\hspace{0pt}}p{#1}}
\newcolumntype{R}[1]{>{\raggedleft\hspace{6pt}}p{#1}}

\newcommand{\Id}{\mathbb{I}}

\newcommand{\D}{{\mathcal D}^{(ov)}}
\newcommand{\bEqn}{\begin{align}}
\newcommand{\eEqn}{\end{align}}

\newcommand{\del}[1]{\partial_{#1}}
\newcommand{\delCov}[1]{\partial^{#1}}

\newcommand{\half}{\frac{1}{2}}

\newcommand{\G}{\mathcal{G}}

\newcommand{\gfive}{\gamma_5}

\newcommand{\Z}{\mathbb{Z}}

\newcommand{\GEV}[1]{$#1$\nopagebreak[4]\nolinebreak[4] GeV}

\newcommand{\lb}{\left ( }
\newcommand{\rb}{\right )}

\newcommand{\br}[1]{\left ( #1 \right )}
\newcommand{\BR}[1]{\left \{ #1 \right \}}

\newcommand{\myCTable}[5]{
   \ctable
      [  cap = {#1},
         caption = {\small #2},
         label = {#5},
         pos=ht,
         doinside={\small }
      ]
      {#3}
      {} 
      { \FL
         #4
      }
}

\newcommand{\fig}[5] {
   \begin{figure}[ht]
   \centering
   \includegraphics[width=#4]{#5}
   \caption[#2]{#1}
   \label{#3}
   \end{figure}
}
\newcommand{\figOneOne}[5] {
   \begin{figure}[ht]
   \centering
   \begin{tabular}{c}
      \includegraphics[height=#4]{#5}
   \end{tabular}
   \caption[#2]{\small #1}
   \label{#3}
   \end{figure}
}

\newcommand{\figOneTwo}[6] {
   \begin{figure}[ht]
   \centering
   \begin{tabular}{cc}
      \includegraphics[height=#4]{#5} &
      \includegraphics[height=#4]{#6}   \\
      \small\hspace{1.0cm}(a) & \small\hspace{1.0cm}(b) \\
   \end{tabular}
   \caption[#2]{\small #1}
   \label{#3}
   \end{figure}
}

\newcommand{\figOneThree}[7] {
   \begin{figure}[t]
   \centering
   \begin{tabular}{ccc}
      \includegraphics[height=#4]{#5} & \hspace{-0.7cm}
      \includegraphics[height=#4]{#6} & \hspace{-0.7cm}
      \includegraphics[height=#4]{#7}   \\
      \small\hspace{0.75cm}(a) & \small\hspace{0.2cm}(b) & \small\hspace{0.2cm}(c)\\
   \end{tabular}
   \caption[#2]{\small #1}
   \label{#3}
   \end{figure}
}

\newcommand{\eqn}[2]{
   \begin{align}
      #1
      \label{#2}
   \end{align}
}
\renewcommand{\eqn}[2] {
   \begin{multline}
   #2
   \label{#1}
   \end{multline}
}
\newcommand{\eq}[1]{
   \begin{equation*}
      \begin{aligned}
      #1
      \end{aligned}
   \end{equation*}
}
\newcommand{\EQ}[1]{
   \begin{align}
#1
   \end{align}
}

\newcommand{\VEC}[1]{
   \begin{pmatrix}
#1
   \end{pmatrix}
}
\graphicspath{{images/}}

\begin{document}

\begin{frontmatter}



\title{\hfill \normalsize{DESY 11-218}\vspace{1cm}\\
\Large The Higgs boson resonance width from a chiral Higgs-Yukawa model on the lattice}


\author{Philipp Gerhold\fnref{a,b}\footnote{present address: d-fine GmbH, Opernplatz 2, 60313 Frankfurt}}
\author{Karl Jansen\fnref{a}}
\author{Jim Kallarackal\fnref{a,b}\footnote{present address: OakLabs GmbH, Am M\"uhlenberg 11, D-14476 Potsdam}}

\address[a]{NIC, DESY, Platanenallee 6, D-15738 Zeuthen, Germany }
\address[b]{Humboldt-Universit\"at zu Berlin, Institut f\"ur Physik, Newtonstr. 15, \par D-12489 Berlin, Germany }

\begin{abstract}
The Higgs boson is a central part of the electroweak theory
and is crucial to generate masses for quarks, leptons and the weak gauge bosons.
We use a 4-dimensional Euclidean lattice formulation
of the Higgs-Yukawa sector of the electroweak model
to compute physical quantities in the path integral approach
which is evaluated by means of Monte Carlo simulations thus allowing for
fully non perturbative calculations.
The chiral symmetry of the
model is incorporated by using the Neuberger overlap Dirac operator.
The here considered Higgs-Yukawa model
does not involve the weak gauge bosons and furthermore, only a degenerate
doublet of top- and bottom quarks are incorporated.
The goal of this work is to study the resonance properties of the Higgs boson
and its sensitivity to the strength of the quartic self coupling.
%
\end{abstract}


\end{frontmatter}


\section{Introduction}\label{sec:introduction}
The model under consideration is the pure
Higgs-Yukawa sector of the electroweak standard model
formulated on a 4-dimensional Euclidean space-time lattice.
The field content of the model consists
of a complex scalar Higgs doublet and a fermion
doublet coupled in a chirally invariant way to the scalar sector.
Our lattice model closely resembles the weak interaction of the
continuum standard model in the limit where the gauge fields are
switched off.
The chiral nature of the weak interaction
is realized in our lattice setup with the help of the Neuberger overlap Dirac
operator \cite{Neuberger:1998wv} which allows to incorporate an exact chiral symmetry
on the lattice \cite{Luscher:1998pqa}.

Within the past years renewed efforts have been undertaken to
investigate the lower and upper Higgs boson mass bounds
within this model in the framework of
lattice field theory
\cite{Gerhold:2010bh,Gerhold:2009ub,Fodor:2007fn}, see these
references also for accounts of earlier lattice investigations
of Higgs boson mass bounds.
However, the decay width of the Higgs boson was not taken into
account in refs.~\cite{Gerhold:2010bh,Gerhold:2009ub}
assuming that its effect on the Higgs boson mass is small. The first
main aim of
this work is to treat the Higgs boson as a true resonance and to
investigate the unstable nature of the Higgs boson from first
principles. In this way, the assumption of the preceding work
\cite{Gerhold:2010bh,Gerhold:2009ub}
can be tested and it can be investigated, whether the
Higgs boson mass bounds are affected.
The second goal is to determine the Higgs boson width as a function
of
the quartic self-interaction of the Higgs fields, including large values
of the coupling.
In particular, we want to investigate, whether in the decay channel
of the Higgs boson into massive vector bosons the Higgs boson becomes
a broad resonance at large couplings, or whether the width is still rather narrow.

The strategy to determine the Higgs boson resonance parameters
is the
analysis of the scattering phases. The
computation of the scattering phase within lattice field
theory is in turn based on the determination of the volume dependence of
energy eigenvalues \cite{Luscher:1990ux}.
An extension
of the finite size method of ref.~\cite{Luscher:1990ux},
which was formulated in the center of mass frame,  has been proposed in
\cite{Rummukainen:1995vs} and is based on the analysis of the energy
levels within a moving frame. The method is complementary to the
original method and allows to obtain more energy levels
on the same underlying field configurations. We employ
this method here to reduce the errors on our extraction of the
scattering width and the resonance mass, which we are then able
to determine at the $O(10\%)$ level.

\section{The model}
\label{sec:model}
The Higgs-Yukawa model is defined by the Lagrangian and the corresponding generating
functional for the Green functions of the theory.
With regard to the lattice formulation of the model, the Euclidean version of the model will
be considered here.
The particle content contains the scalar sector and the heaviest quark doublet
consisting of the top and the bottom quark. Due to the neglection of gauge bosons
the Lagrangian exhibits a {\em global} $SU_W(2) \times U_Y(1)$ inner symmetry rather than
a local (gauge) symmetry.
The Euclidean action is given by
\eqn{eqn:chap_model_cont_lagrangian}{
   L_{E}^{HY} = \half \lb \del{\mu} \varphi \rb^{\dagger} \cdot
                  \lb \delCov{\mu} \varphi \rb
                  + \half m^2 \varphi^{\dagger} \cdot \varphi
                  + \lambda \lb \varphi^{\dagger} \cdot \varphi \rb^{2} \\
               \qquad + \overline{t} \, \slashed{D} \,t + \overline{b}\, \slashed{D}\, b
                  + y_b \begin{pmatrix}\overline{t}\\ \overline b\end{pmatrix}_{L}^{T}\hspace{-0.3cm}\cdot
                  \varphi \; b_{R} +
                  y_t \begin{pmatrix}\overline{t}\\ \overline b\end{pmatrix}_{L}^{T}\hspace{-0.3cm}\cdot
                     \tilde \varphi \; t_{R} \quad + h.c. .
}
Here, $m^2$, $\lambda$ and $y_{t(b)}$ denote the bare scaler mass, the bare quartic coupling
and the bare top (bottom) Yukawa coupling, respectively.
It is common in lattice field theory to rewrite the scalar sector by rescaling the fields
with a factor $\sqrt{2 \kappa}$ and furthermore, the scalar doublet can be expressed as
a quarternion
\EQ{
\phi &:=  \VEC{\tilde \varphi_1 & \varphi_1 \\
            \tilde \varphi_2 & \varphi_2} =: \phi^0 \Id - i \sigma^j \phi^j,
            \quad j \in \{1\dots 3\}, \quad \phi^\mu \in \mathbb{R}.
\label{eqn:scalar_fields}
}
The rescaled fields are:
\eq{
   \Phi^{\mu} &:= \frac{1}{\sqrt{2 \kappa}} \; \phi^{\mu}.
}
The scalar fields in the usual notation of eq.~\eqref{eqn:chap_model_cont_lagrangian}
can be recovered by identifying
\eq{
\VEC{\varphi_1\\\varphi_2}= \sqrt{2\kappa} \VEC{
   \Phi_x^2 + i\Phi_x^1\\
   \Phi_x^0-i\Phi_x^3},
}
where $\varphi_1,\varphi_2$ are the original, complex valued scalar fields.

In 1982 Ginsparg and Wilson \cite{Ginsparg:1981bj} proposed a relation which
defines a class of lattice Dirac operators and which is since then known as the
Ginsparg-Wilson relation
\EQ{
   \gamma_5 D + D \gamma_5 &= a D \gamma_5 R D \label{eqn:Ginsparg-Wilson}.
}
Here $a$ is the lattice spacing and $R$ is a positive constant.
The Ginsparg-Wilson relation can be utilized to construct a lattice modified chiral
symmetry which recovers the desired continuum chiral
symmetry in the limit $a \rightarrow 0$
\cite{Luscher:1998pqa}.
The modified lattice projectors for the left and right handed components are given by
\eq{
   \hat{P}_{\pm} &:= \half \br{1 \pm \hat{\gamma}_5},
      & \hat{\gamma}_5  &:= \gfive \br{1-aRD}.
}

The modified projectors are used to define the chiral components of the spinor fields on the lattice
\eq{
   \psi_R &= \hat P_{+} \psi, &\qquad \psi_L &= \hat P_{-} \psi, \\
   \overline{\psi}_R &= \overline{\psi} P_{-}, &\qquad \overline{\psi}_L &= \overline{\psi} P_{+}.
}

As in the continuum theory the free part of the fermion Lagrangian can be written as a
sum of
the left and right handed lattice spinor fields
\eq{
   \overline{\psi} D \psi &= \overline{\psi}_L D \psi_L + \overline{\psi}_R D \psi_R
}
The modified projectors allow to formulate a theory of
continuum-like left- and right handed
fermions on the lattice for any non-vanishing value of the lattice spacing $a$.

In order to specify the lattice action, a Ginsparg-Wilson type Dirac operator has to be introduced.
The here presented results are based on the Neuberger overlap operator \cite{Neuberger:1998wv}, which
satisfies  the Ginsparg-Wilson relation and is given by
\EQ{\label{eq:chap_model-neuberger_operator}
\D &= \frac{\rho}{a} \BR{1+\frac{ A}{\sqrt{ A^\dagger  A}} },
   \quad A = D^{W} - \frac{\rho}{a}, \quad 0 < \rho < 2r
}
$\rho$ is chosen to be $1$ in this work and
$D^{W}$ denotes the Wilson Dirac operator.
In \cite{Hernandez:1998et} it has been demonstrated that the
Neuberger overlap operator is, despite its complicated form, a local operator.

The full Euclidean discretized action is given by
\eqn{eqn:action}{
S = -\kappa \sum_{x,\mu} \Phi_x^{\dagger} \br{\Phi_{x+\mu} + \Phi_{x-\mu}}
   + \sum_{x} \Phi^{\dagger}_x\Phi_x + \hat\lambda \, \sum_{x} \left(\Phi^{\dagger}_x\Phi_x - N_f \right)^2\\
   +\sum\limits_{x,y} \VEC{\overline{t}_x^{\alpha}\\\overline{b}_x^{\alpha}} \BR{\Id_2 D_{x,y}^{\alpha \beta} +
      \hat y \br{P_-\phi \hat P_{-} + P_+\phi^\dagger \hat P_+ }_{x,y}^{\alpha \beta}  } \VEC{{t}_y^{\beta}\\{b}_y^{\beta}}.
}

In the following, the lattice spacing $a$ is set to unity.
It can be given a physical value by identifying the vacuum expection value
on the lattice, $av_{\rm phys}$ with the measured value of $v_{\rm phys}=246$ GeV.
As mentioned before, the couplings are scaled by the $\kappa$ parameter. The  parameterization
of the Lagrangian in \eqref{eqn:chap_model_cont_lagrangian}
can by recovered with the identities:
\eq{
\lambda&=\frac{\hat\lambda}{4\kappa^2}, &
   m_0^2 &= \frac{1 - 2N_f\hat\lambda-8\kappa}{\kappa}, &
      y_{t,b} = \frac{\hat y_{t,b}}{\sqrt{2\kappa}}.
}

The here considered Higgs-Yukawa model has already been studied
extensively. In refs.~\cite{Gerhold:2007gx,Gerhold:2007yb}
the phase structure has been determined, lower and upper Higgs boson mass bounds
have been obtained in \cite{Gerhold:2010bh,Gerhold:2009ub} and also
an extension to a possible fourth quark generation
has been discussed in \cite{Gerhold:2010wv}. Many details of the model, its
numerical implementation and algortihmic aspects can be found
in ref.~\cite{Gerhold:2010wy}.


\section{Resonances from lattice simulations}
\label{sec:theory}

The method to compute the resonance parameters in finite volume
was proposed in \cite{Luscher:1990ux}. It has been
successfully employed for the case of the Higgs boson mass
in the pure $\lambda \phi^4$ theory
\cite{Gockeler:1994rx}; here we want to extend
the method to the Higgs-Yukawa model as considered in this work.

Let us start with the optical theorem which
states that the total cross section in the elastic region of
forward scattering is given by
\EQ{\label{eqn:total_cross_section}
  \sigma_{tot} &= \frac{8 \pi}{q \sqrt{s}} \sum_{j=0}^{\infty} \br{2j + 1} \Im \br{\mathcal{T}_{ii}^{j}}
}
$q$ denotes the centre of mass momentum and $s$ is the centre of mass
energy of the two incomming particles with four momentum $p_a$ and $p_b$,
$s = \br{p_a + p_b}^2$. $\mathcal{T}_{ii}^{j}$ is the spherical decomposition of the
interacting part $T$ of the S-Matrix
\eq{
S=\Id + i T.
}
The total cross section exhibits a peak near a resonance which can be well
parameterized with a
Breit-Wigner function containing the resonance mass $m$ and the width $\Gamma$,
\EQ{\label{eqn:breitwigner_scattering_phase}
  \sigma_{tot} &= \Big | \frac{1}{p^2 - m^2 + im \Gamma} \Big |^2 =
    \frac{8 \pi}{q \sqrt{s}} \sum_{j=0}^{\infty} \br{2j + 1} 2 \sin^2 \delta_j (s)\; ,
}
where we have also expressed
the total cross section
in eq.~\eqref{eqn:total_cross_section} in terms of the
scattering phases $\delta_j$.

In order to develop a general method to extract the scattering phase from the
energy levels in a finite box, it is sufficient to
investigate the problem in non-relativistic quantum mechanics. The
non-relativistic result can then be transferred to the case of quantum field theory.
This remarkable result has been demonstrated in
refs.~\cite{Luscher:1990ck,Luscher:1986pf}.

As a result of these works the computation of the scattering phase $\delta_0$
in which we are interested in, proceeds as follows. The first step is to
compute the energies of the two-particle massive Goldstone system, $W=2\sqrt{m^2 + k^2}$,
from which
the value of $k$ can be extracted. Knowing $k$, the scattering phase $\delta_0$
is obtained from
\eq{
  \tan \delta_0 (k) &=  \frac{\pi^{\frac{3}{2}}q}{\mathcal{Z}_{00} \br{1; q^2}}, \quad
  q = \frac{k L}{2 \pi}, \\
  \mathcal{Z}_{00} \br{1; q^2} &= \sum_{\vec{n} \in \Z^3} \frac{1}{\sqrt{4 \pi}}  \frac{1}{n^2 - q^2}.
}

The analysis of scattering phases has been extended to moving frames in
\cite{Rummukainen:1995vs,Feng:2010es,Feng:2011ah}, where one of the
two particles is at rest.
The method is complementary to the centre of mass frame and allows to compute
more data for the scattering phases from the same configurations. The obtained
two particle energies have to be translated back to the centre of mass frame.
In general, the choice of a moving frame implies that one has to consider
irreducible representations of a subgroup of the cubic group.
The remaining symmetry depends on the selected directions of the moving frame.
The here used irreducible representations are given in
\cite{Rummukainen:1995vs}. The modification of the relation between
the two particle energy in the moving frame and the scattering phase is
given by
\eq{
  \tan \delta_0 (q) &= \frac{\gamma q \pi^{\frac{3}{2}}} {\mathcal{Z}_{00}^{d} (1; q^2)},
  \qquad q=\frac{p^* L}{2 \pi}\; ,
}
where $p^*$ denotes the momentum which has been transferred back to the centre of mass
frame by a Lorentz boost. The modified Zeta function is defined by
\eq{
  \mathcal{Z}_{00}^d (s;q^2) &= \frac{1}{\sqrt{4 \pi}} \sum_{r \in P_d} \frac{1}{\br{r^2-q^2}^2}, \\
  P_d &:=
  \BR{\vec{r} \in \mathbb{R}^3 | \vec{\gamma}^{-1}\br{\vec{n} + \half\vec{d}}, \vec{n} \in \Z^3 }.
}
The vector $\vec{d}$ is related to the total momentum of the moving frame $\vec{P}$.
$\gamma$ is the usual Lorentz factor.
Below are some definitions which are needed to compute the
modified Zeta function
\eq{
  \vec{P}&:= \frac{2\pi}{L}\, \vec{d} \\
  \gamma &= \frac{1}{\sqrt{1-v^2}}, \quad \vec{v} = \frac{\vec{P}}{W_L}\\
  \vec{\gamma}^{-1} \vec{n}&= \gamma^{-1} \vec{n}_{||} + \vec{n}_{\bot}.
}
$W_L$ denotes the two particle energy in the moving frame which can be
computed with time slice correlators. The corresponding observables are
defined in the next section.
$\vec{n}_{||}$ and $\vec{n}_{\bot}$ is a decomposition of the vector $\vec{n}$
in its parallel and perpendicular parts with respect to the centre of mass
velocity $\vec{v}$
\eq{
  \vec{n}_{||} &:= \frac{\br{\vec{n} \cdot\vec{v}} \vec{v}}{v^2}\\
  \vec{n}_{\bot} &:= \vec{n} - \vec{n}_{||}.
}

\section{Numerical results}
\label{sec:numerics}

The Higgs boson decays dominantly to any even number of
Goldstone bosons, if kinematically allowed. The physical set-up chosen here
allows always for such
a decay, i.e. we were always working in
the elastic scattering region.
In order to achieve this situation, external currents were introduced,
providing appropriate masses to the else mass-less Goldstone bosons.
The bare parameters used in our simulations are given
in table~\ref{tab:chap_sm_physics:bare_param}.
The corresponding physical value of the cut-off, the Higgs boson
propagator mass, the Goldstone boson mass and the
obtained top quark mass are summarized in table \ref{tab:chap_sm_physics:sim_param}.

\myCTable
  {Simulation parameters for the scattering phases.}
  {The table summarizes the bare parameters for the Monte Carlo simulations
   which were performed in order to determine the scattering phases. $J$ denotes an external
   current which is coupled to one component
   of the scalar field. $J$ is adjusted such that the
   the ratio of the Higgs boson mass $M_H^p$ and
   and the Goldstone boson mass
   $M_G^p$ is approximately $3$ (see table~\ref{tab:chap_sm_physics:sim_param}),
   which ensures that the Higgs boson
   decays into any even number of Goldstone bosons.
  }
  {p{0.2\textwidth}p{0.2\textwidth}p{0.2\textwidth}p{0.2\textwidth}}
  {
   $\hat \lambda$ & $\kappa$ & $\hat y$ & $J$ \ML
   $0.01$ & $0.1295$ & $0.36274$ & $0.001$ \\
   $1.0$ & $0.2445$ & $0.49798$    & $0.002$\\
   $\infty$ & $0.3020$ & $0.57390$ & $0.002$\LL
  }
  {tab:chap_sm_physics:bare_param}
\myCTable
  {Physical parameters for the scattering phases.}
  {The table summarizes the obtained masses in lattice units for the bare
  parameters given in table \ref{tab:chap_sm_physics:bare_param}.
   The Higgs and Goldstone boson
   masses, $M_H^p$ and $M_G^p$ respectively, are extracted from the propagator
   {\em neglecting} the width.
   The quark mass is computed from the corresponding
   time slice correlator and the last columns show
   the renormalized vacuum expectation value $v_R$ and the cut-off ($\Lambda$).
   For the smaller lattices we have typically used 20000 and for the
   largest lattice typically 10000 configurations. The autocorrelation
   times were about one in all cases, see also \cite{Gerhold:2010wy}.
   Note that, due to a technical problem,
   for the determination of $M_H^p$ at ${\hat\lambda=1.0}$ we have used
   a significantly smaller statistics than at the other simulation points.
  }
  {p{0.12\textwidth}p{0.12\textwidth}p{0.12\textwidth}p{0.12\textwidth}p{0.12\textwidth}p{0.12\textwidth}}
  {
   $\hat \lambda$ &
      $a M_H^p$ & $a M_G^p$ & $m_t$ \scriptsize [GeV] & $a v_R$ & $\Lambda$ \scriptsize [GeV]\ML
      $0.01$ &
         $0.278(1)$ & $0.085(2)$ & $174(1)$ & $0.2786(3)$ & $883(1)$\\
      $1.0$ &
         $0.386(28)$ & $0.133(4)$ & $179(2)$& $0.1637(5)$ &$1503(5)$ \\
      $\infty$ &
         $0.405(4)$ & $0.129(1)$ & $178(1)$ & $0.1539(2)$ & $1598(2)$ \LL
  }
  {tab:chap_sm_physics:sim_param}

The Goldstone theorem ensures that the Goldstone bosons are massless. Due to an
external current $J$ which couples to one of the components
of the scalar fields in the complex
$SU_W(2)$ doublet, the symmetry is broken explicitly in the Lagrangian. The
Goldstone bosons acquire a mass and they form a vector under cubic
rotations. The magnitude of the current $J$ is chosen such that the
ratio of the Higgs boson mass $M_H^p$ to the Goldstone boson mass $M_G^p$ is
roughly $3$. Here and below the superscript $p$ in $M_H^p$ and $M_G^p$
denotes that the mass was extracted from the analysis of the momentum space
propagator. This numerically computed propagator was
then fitted to a formula motivated from perturbation theory and the real part
of the complex pole of this fit function has been identified with
the Higgs boson mass, see
refs.~\cite{Gerhold:2010bh,Gerhold:2009ub,Gerhold:2010wv,Gerhold:2010wy}.

The Higgs field is a singlet under cubic rotations and
transforms as elements in the $A_1^+$ representation.
The two particle
energies discussed in the previous section are constructed from the
two particle Goldstone singlet as it has the same quantum
numbers as the Higgs boson.

The analysis of the resonance parameters involves several lattice volumes with identical
bare parameters in order to compute the momentum dependence of the scattering phase.
As shown in table~\ref{tab:chap_sm_physics:bare_param},
there are three distinct set of simulation parameters which shall be characterized
with the value of the bare quartic coupling ($\hat{\lambda} \in \BR{0.01, 1.0, \infty}$).
For each of the three values of the quartic coupling the simulations were
performed on lattice volumes up to $40^4$.
In the following the two particle energies of the two Goldstone boson states will be discussed.
Once these energy levels are known, the unstable nature of the Higgs boson
can be studied by the method described in the preceeding chapter.

The Goldstone bosons are stable particles such that their ground state energy
can be calculated from the two point time correlation function. The concept
of time correlators is widely used in lattice field theory and there are reliable
techniques to extract mass eigenvalues from such correlators.
The method of choice in this work, is the analysis of the correlation matrix
\cite{Luscher:1990ck,Blossier:2009kd}. The
correlation matrix built from an operator $O_{\alpha} (t)$ is given by
\eq{
  C_{\alpha \beta} (\Delta t) &:= \frac{1}{L_t} \sum _{|t-t'|=\Delta t}
        \Braket{O_{\alpha} (t) O_{\beta} (t') }_c\; ,
}
where $L_t$ is the temporal size of the lattice. Throughout this chapter the
temporal extent will be $L_t=40$. The subscript $c$ denotes that the
disconnected part of the correlator has been subtracted. It has been shown in
\cite{Luscher:1990ck} that the eigenvalues of the correlation matrix
decay exponentially with rising time separation $\Delta t$.
In the following
the operators which contribute to the two Goldstone system are collected.

The definition of the observables in the centre of mass frame is given by
\eq{
\mathcal{O}_{0} (t) &:=
        \tilde H (\vec{0}, t)\\
\mathcal{O}_{1} (t) &:=
    \frac{1}{\sqrt{3}} \frac{1}{|Q_1|}
      \sum_{\vec{n} \in Q_{1}}
        \tilde \G^T (\vec{n}, t) \tilde \G (-\vec{n}, t)\\
    & Q_1 = \BR{\vec{n} \in \Z^3| n^2 = 0}, \qquad |Q_1| = 1\\
\mathcal{O}_{2} (t) &:=
    \frac{1}{\sqrt{3}}  \frac{1}{|Q_2|}
      \sum_{\vec{n} \in Q_{2}}
        \tilde \G^T (\vec{n}, t) \tilde \G (-\vec{n}, t)\\
    & Q_2 = \BR{\vec{n} \in \Z^3| n^2 = 1}, \qquad |Q_2| = 6\\
\mathcal{O}_{3} (t) &:=
    \frac{1}{\sqrt{3}}  \frac{1}{|Q_3|}
      \sum_{\vec{n} \in Q_{3}}
        \tilde \G^T (\vec{n}, t) \tilde \G (-\vec{n}, t)\\
    & Q_3 = \BR{\vec{n} \in \Z^3| n^2 = 2}, \qquad |Q_3| = 12.
}
The correlation matrix is thus a $4 \times 4$ matrix.
$\tilde H (\vec{0}, t)$ and $\tilde \G (\vec{n}, t)$ denote the
standard, zero momentum
projected
Higgs and Goldstone boson interpolating fields $H(x)$, $\G(x)$,
see e.g. \cite{Gockeler:1994rx,Gerhold:2010wy}.

In order to collect more data for the scattering phases, the modification of the method
to a moving frame was analysed as well. The moving frame is characterised by a constant
vector $\vec{d}$ which indicates the momentum of the frame. The observables for the
moving frame are  constructed such that one of the Goldstone bosons is at rest while the
other can take any momentum allowed on the lattice. The selection of a constant vector
$\vec{d}$ breaks the cubic symmetry and thus special care is needed while constructing
the observables. Fortunately, it turns out that the $A_1^+$ sector does not need
much modification and explicit relations are given in \cite{Rummukainen:1995vs}.
The lowest energy eigenstates are associated to the lowest possible relative
momentum and thus only moving frames with momentum $\vec{d}= (0,0,1)$ and
permutations thereof will be considered.  The observables
are ($\vec{e}_i$ is the unit three-vector in direction~$i$)
\eq{
  \vec{d}_i = \vec{e}_i\\
  \mathcal{O}_{\vec{d}_i, 0} (t) &:=
    \tilde H (\vec{d}, t)\\
  \mathcal{O}_{\vec{d}_i, 1} (t) &:=
    \tilde \G^T (\vec{d}_i, t) \tilde \G (\vec{0}, t)\\
  \mathcal{O}_{\vec{d}_i, 2} (\Delta t) &:=
    \frac{1}{4}
      \sum_{\vec{n} \in Q_{d_i, 2}} \tilde \G^T (\vec{n} + \vec{d}_i, t) \tilde \G (-\vec{n}, t)\\
      & Q_{d_i, 2} = \BR{\vec{n} \in \Z | \vec{n}\cdot \vec{d}_i = 0, n^2 = 1} \\
  \mathcal{O}_{\vec{d}_i, 3} (t) &:=
        \tilde \G^T (2\vec{d}_i, t) \tilde \G (-\vec{d}_i, t).\\
}
The correlation matrix in the moving frame is then given by
\eq{
  C_{\alpha \beta} (\Delta t) &:= \frac{1}{3} \frac{1}{L_t}\sum_{|t-t'|=\Delta t}\;
      \sum_{i=1}^{3}
      \Braket{{O}_{\vec{d}_i, \alpha} (t) \; {O}_{\vec{d}_i, \beta} (t')}_c \\
      & \alpha, \beta \in \BR{1,2,3}.
}
The energy levels obtained from the moving frame
are connected to energy levels in the corresponding centre of mass frame by
Lorentz transformation.

Fig.~\ref{fig:chap_sm_physics-final_result_cross sections} shows the obtained
cross sections and Fig.~\ref{fig:chap_sm_physics-final_result_phases}
the obtained scattering phases for the three different physical situations
(different values of $\hat{\lambda}$) we have used.
The cross sections are plotted against the energy $W_k$ while
the scattering phase, which take values in the interval $[0,\pi]$, are plotted against
the momentum $k$. If the scattering phase $\delta (k)$ passes through $\frac{\pi}{2}$
it indicates the existence of a resonance. Hence, all three setups involve an unstable Higgs
boson.
The cross section can be decomposed into spherical harmonics and is parametrised by
the scattering phase
\EQ{\label{eqn:scattering_phase-cross_section}
  \sigma (k) = \frac{4 \pi}{k^2} \sum_{j = 0}^{\infty} \br{2j + 1} \sin^2 \br{\delta_j (k)}
}
which becomes
\eq{\label{eqn:scattering_phase-cross_section_simple}
    \sigma (k) \approx \frac{4 \pi}{k^2} \sin^2 \br{\delta_0 (k)},
}
when the contribution of the higher angular momenta $j > 0$ are neglected.
We use a fit function to describe our data for $\sigma (k)$ which is motivated
by a
Breit-Wigner form and allows us to determine the
resonance mass $M_H$ and the
width $\Gamma_H$:
\EQ{
  f(k) &:= 16 \pi \frac{M_H^2 \Gamma_H^2}
    {\br{M_H^2 - 4 m_G^2} \br{\br{W_k^2 - M_H^2}^2 + M_H^2 \Gamma_H^2}}.
    \label{eqn:fitfunction}
}
In fig.~\ref{fig:chap_sm_physics-final_result_cross sections},
the solid curve shows the result of the fit for the cross section and
in
fig.~\ref{fig:chap_sm_physics-final_result_phases} we show the corresponding
scattering phases. The scattering phase is directly related to the cross section
in equation \eqref{eqn:fitfunction} as can be seen in
equation \eqref{eqn:breitwigner_scattering_phase}. 
Given the scattering phases,
the cross section can be obtained from equation 
\eqref{eqn:scattering_phase-cross_section} setting $j=0$. 

\figOneThree
  {The figure shows the cross-sections as obtained in three different
  physical situations for $\hat \lambda = 0.01$ (a),
  $\hat \lambda =1.0$ (b) and $\hat \lambda = \infty$ (c).
  The cross sections are plotted against the energy $W_k$.
  The red points refer to cross sections obtained from
  the analysis in the centre of mass frame as originally proposed in
  \cite{Luscher:1990ux}. The blue points denote the cross sections computed
  within a moving frame
  \cite{Rummukainen:1995vs}.
  The solid line is a fit to the data using
  eq.~(\ref{eqn:fitfunction}). The computations were performed on
  various lattice volumes
  $L_s^3 \times 40$ where $L_s \in \BR{12,16,18,20,24,32,40}$.}
  {Cross sections.}
  {fig:chap_sm_physics-final_result_cross sections}
  {4.0cm}
  {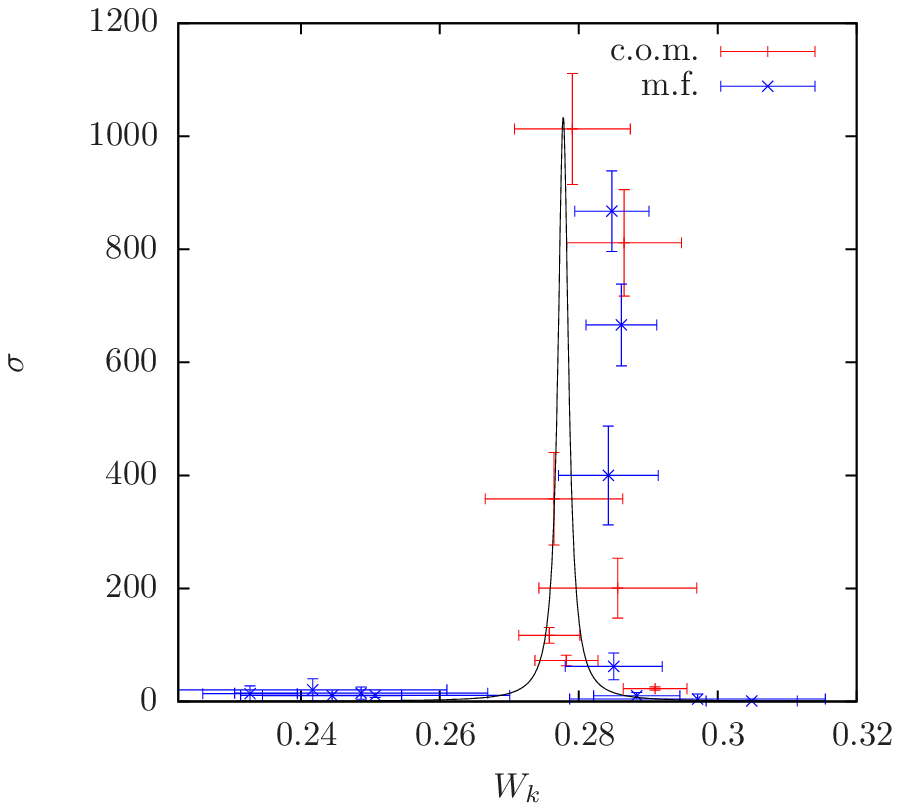}
  {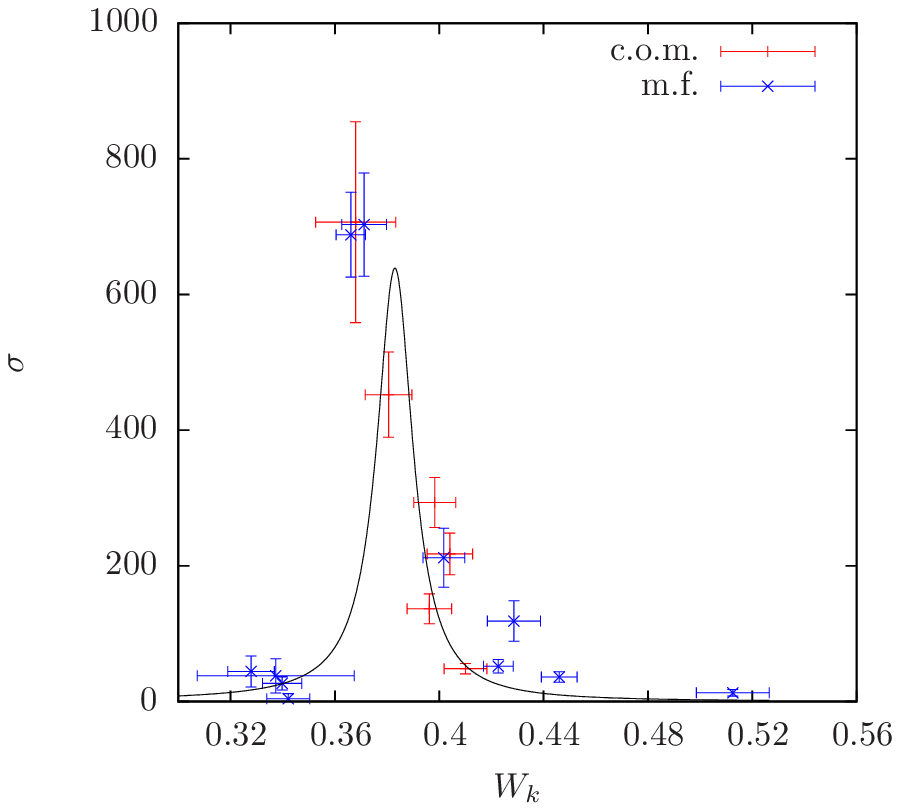}
  {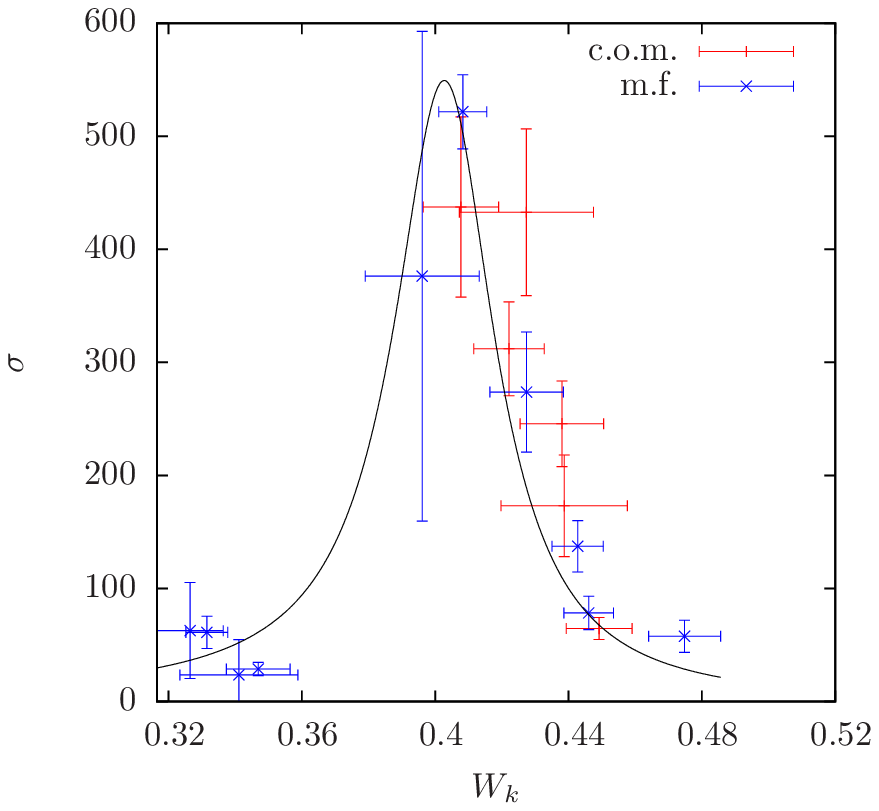}

\figOneThree
  {The figure shows the scattering phases
  that correspond to the
  cross sections shown in fig.~\ref{fig:chap_sm_physics-final_result_cross sections}
  from which we also take the notation.
  The scattering phases are plotted as a function of the momentum $k$.
  The vertical dotted line indicates the inelastic
  threshold. The solid line shows the scattering phase
  obtained from the fit function to the cross sections displayed in
  fig.~\ref{fig:chap_sm_physics-final_result_cross sections}.}
  {Scattering phases.}
  {fig:chap_sm_physics-final_result_phases}
  {4.0cm}
  {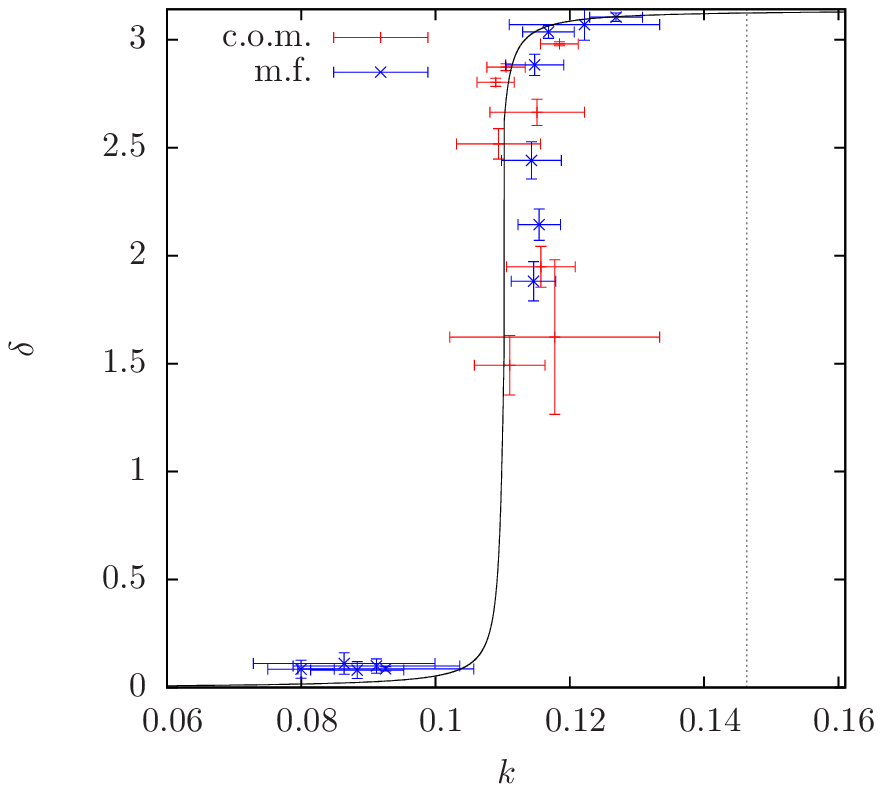}
  {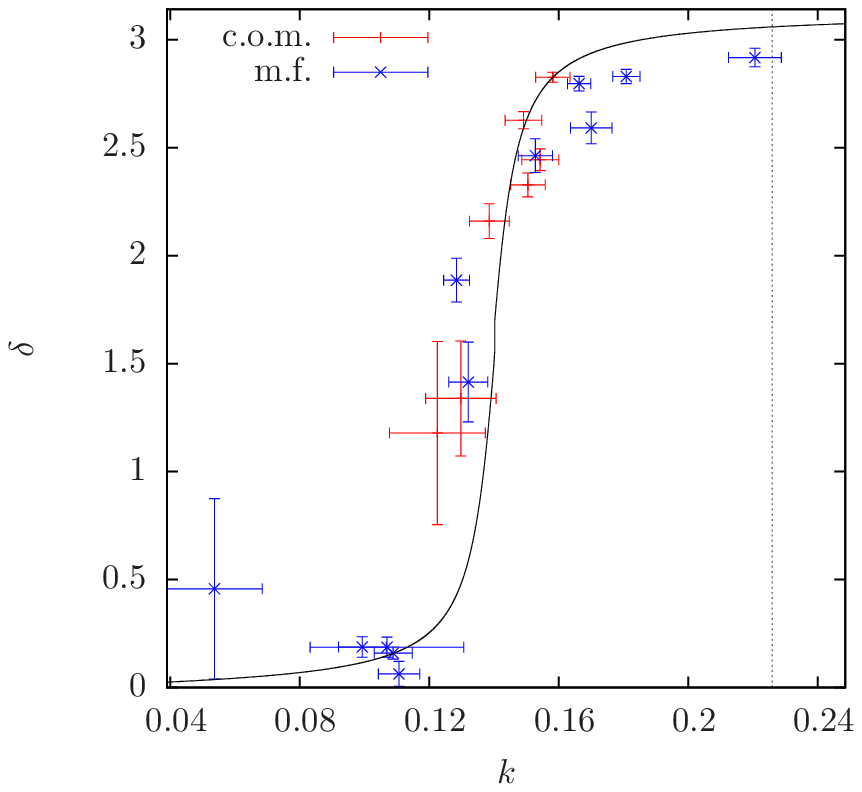}
  {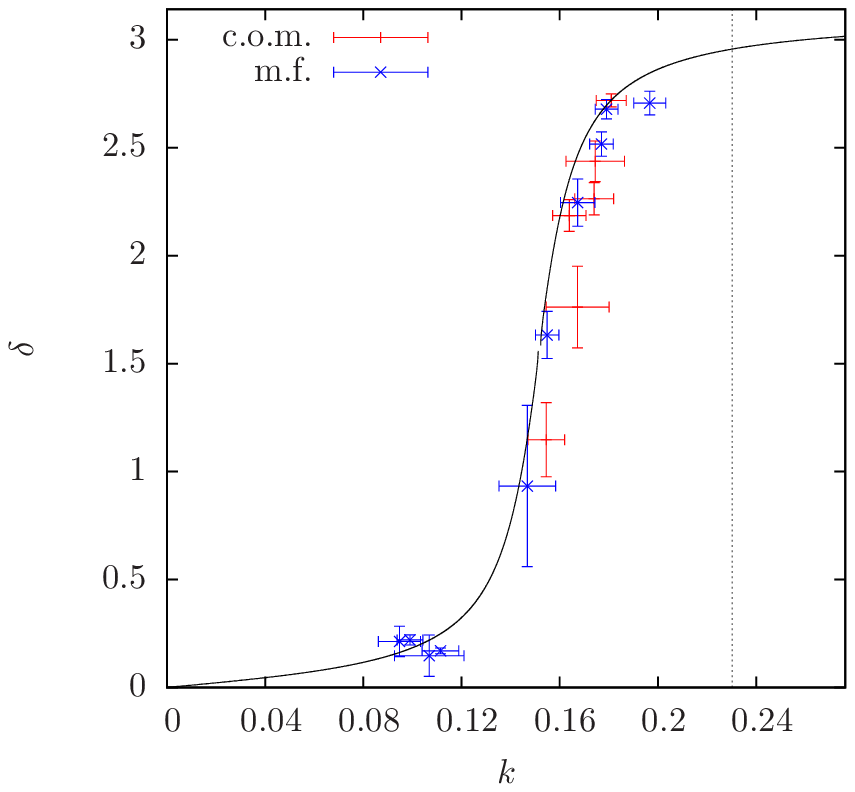}

Finally table \ref{tab:resonance_mass_compare} summarizes the results
for the Higgs boson mass obtained
by different approaches. The physical (resonance) Higgs boson mass
$a M_H$ is compared to the
mass obtained from the Higgs boson propagator $a M_H^p$ and the
energy eigenvalue from
the correlation matrix analysis (GEVP) which corresponds
to the Higgs boson mass.
The results for $a M_H^p$ and the GEVP were obtained
after an extrapolation to infinite volume, see
fig.~\ref{fig:infinite_volhiggs}
for the example of
the infinite volume extrapolation of the eigenvalue from the GEVP.
For completeness, we also show
the finite size behaviour of the renormalized
vacuum expectation value and the top quark mass in fig.~\ref{fig:infinite_vol}.

\myCTable
  {Comparison of the resonance mass with different approaches.}
  {The table summarizes the obtained final results on the resonance mass
  $M_H$ and the resonance width $\Gamma_H$ of the Higgs boson.
  $\hat \lambda$ denotes the bare quartic coupling. $\Lambda$
  is the cut-off of the theory.
  The following two columns display the resonance parameters computed
  from the scattering phases. The $\chi^2$ per degree of freedom
  from the fits to obtain the resonance parameters are
  $\chi^2/d.o.f.=1.3$, $\chi^2/d.o.f.=1.0$ and $\chi^2/d.o.f.=1.2$
  for $\hat \lambda=0.01$, $\hat \lambda=1.0$ and
  $\hat \lambda=\infty$, respectively. $\Gamma_H^{\rm pert}$ is the width obtained from
  perturbation theory where a non vanishing mass for the Goldstone bosons
  has been considered \cite{Luscher:1988uq}.
  Finally the mass extracted from the propagator
  as well as the eigenvalue corresponding to the Higgs boson mass
  computed with the help of the correlation
  matrix is shown.}
  {cccccccc}
  {
  $\hat \lambda$&
   $\Lambda$ [GEV]&
    $a M_H$ &
      $a \Gamma_H$ &
         $a \Gamma_H^{\text{pert}}$ &
            $a M_H^p$ &
               GEVP \ML
  $0.01$   & $883(1)$ & $0.278(3)$ & $0.0018(14)$ & $0.0054(1)$ & $0.278(2)$   & $0.274(4)$ \\
  $1.0$    & $1503(5)$ & $0.383(6) $ & $0.0169(4)$ & $0.036(8)$ & $0.386(28)$ & $0.372(4)$ \\
  $\infty$ & $1598(2)$ & $0.403(6) $ & $0.037(9)$ & $0.052(2)$ & $0.405(4)$   & $0.403(7)$ \LL
  }
  {tab:resonance_mass_compare}

\figOneOne
  {The figure shows the energy eigenvalues corresponding to the Higgs boson
  mass
  obtained by the correlation matrix analysis. The eigenvalues
  are plotted against the inverse squared spatial lattice extent in order to
  perform an extrapolation to infinite volume.
  From top to bottom the data correspond to
  $\hat \lambda=\infty$, $\hat \lambda=1$ and $\hat \lambda=0.01$}
  {Infinite volume extrapolation of GEVP Higgs mass.}
  {fig:infinite_volhiggs}
  {0.5\textwidth}
  {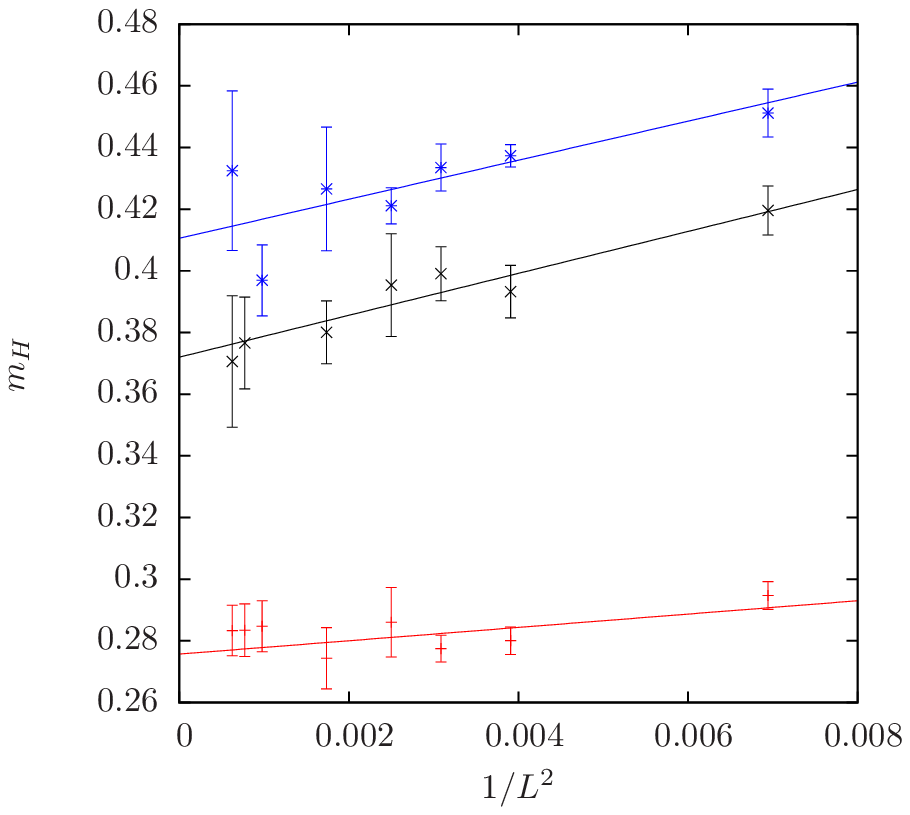}

\figOneTwo{The figure shows the finite size effects of the scalar
vacuum expectation value
  and the fermion mass.
  The top quark mass is computed
  from the fermion time slice correlation function. The figure shows
  also an extrapolation to infinite volume starting from lattice
  volumes of at least $16^3$. From top to bottom the data correspond to
  $\hat \lambda=0.01$, $\hat \lambda=1$ and $\hat \lambda=\infty$}
  {Infinite volume extrapolation.}
  {fig:infinite_vol}
  {5cm}
  {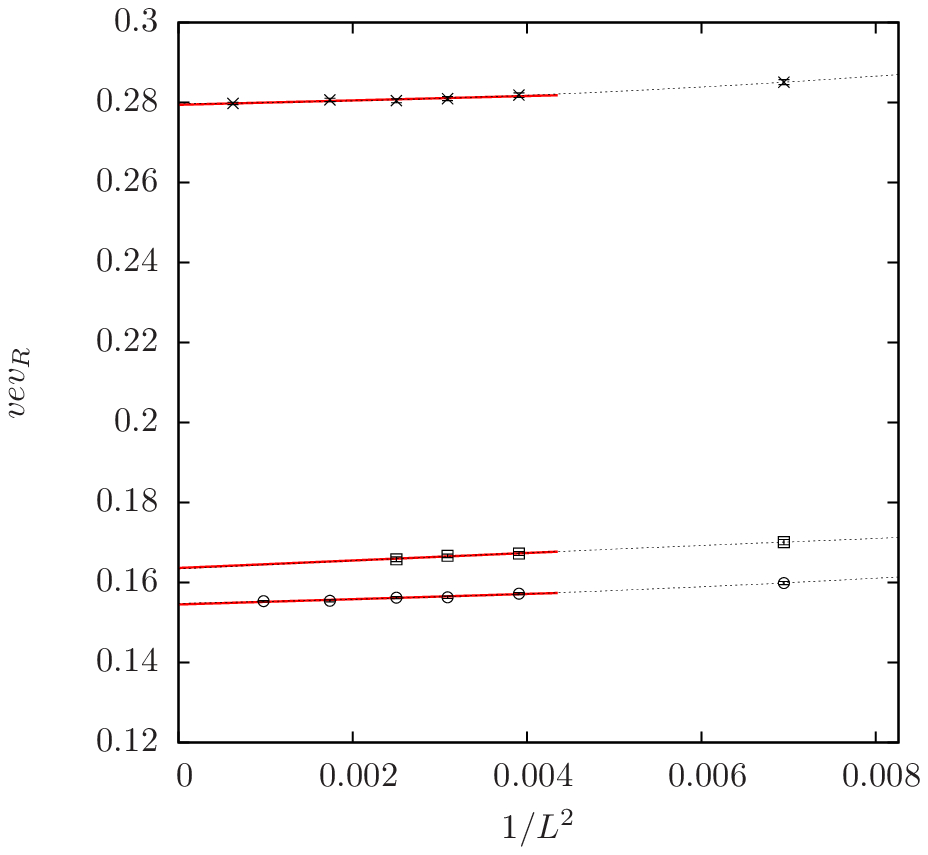}
  {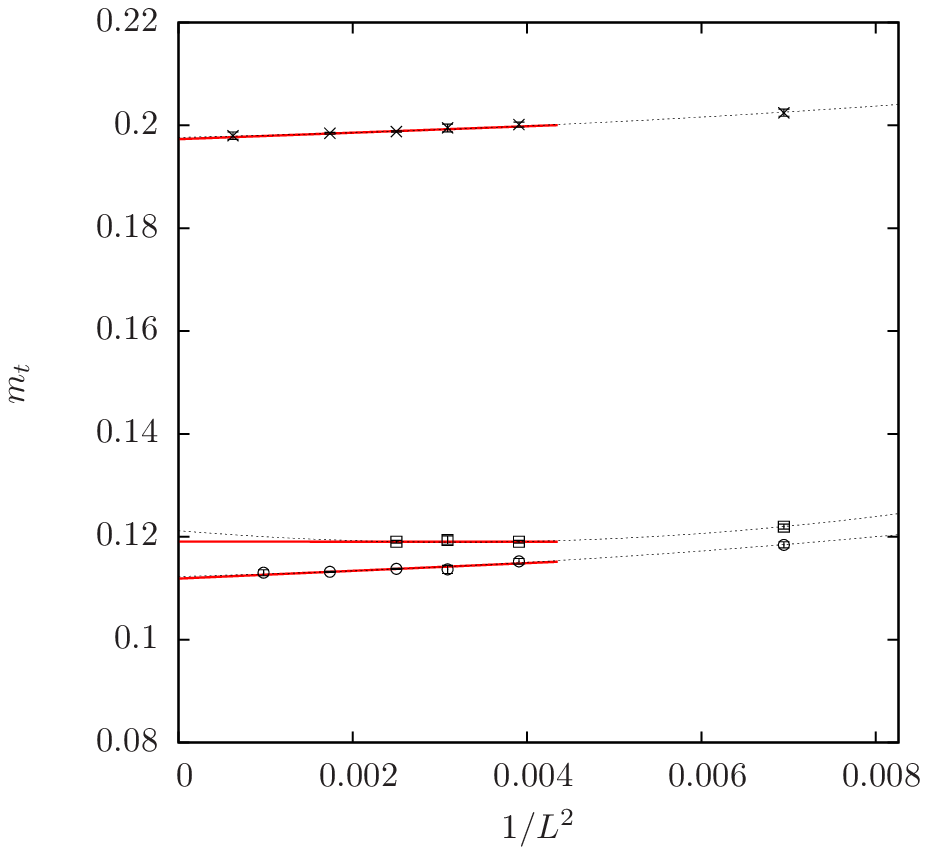}

The simulations at $\hat \lambda=1.0$ and $\hat \lambda =\infty$ belong to
a cut-off of around $1.5$ TeV.
It was necessary to reduce the cut off to \GEV{880} for the smallest
quartic coupling $\hat \lambda=0.01$ in order to meet the resonance condition
($\frac{M_H^p}{m_G} \approx 3$). As one can see from table
\ref{tab:resonance_mass_compare}, the Higgs boson mass is
still well below the cut off.
Especially for the smallest quartic coupling $\hat \lambda=0.01$ the resonance region
is very small ($m_G = 0.09(1) \Rightarrow 0.18 \le W_k \le 0.36$) which in turn
necessitates large lattice volumes in order to obtain energy eigenvalues
which lead to scattering phases near the resonance mass. The plots in
fig.~\ref{fig:chap_sm_physics-final_result_cross sections} and
fig.~\ref{fig:chap_sm_physics-final_result_phases} show that the
analysis of the moving frame can help significantly to extract reliable
results.
\clearpage

\section{Conclusion}\label{sec:conclusion}

In this letter we have investigated the resonance parameters
of the standard model Higgs boson from simulations
on a Euclidean lattice in the limit of the
electroweak theory where the gauge fields are neglected.
In this situation the Higgs boson decays into massive
vector bosons.
The resonance mass and the width of the Higgs boson is
computed using the finite size method proposed in
\cite{Luscher:1990ux}, applicable for the center
of mass frame, combined with the extension
to a moving frame
as proposed in \cite{Rummukainen:1995vs}.
Using both kind of Lorentz frames a large number of scattering phases
could be obtained which allowed us to compute
the Higgs boson resonance mass and width with an accuracy
of $O(1\%)$ for the resonance mass and $O(10\%)$ for the width,
see table~\ref{tab:resonance_mass_compare}.


We have employed three values of the quartic coupling $\hat{\lambda}$ of
the Higgs field ranging from small perturbative values
to $\hat{\lambda}=\infty$ with parameters corresponding to a large renormalized
quartic coupling.
In all three cases the Higgs boson width is not larger than about $10 \%$ with
respect to the resonance mass. Therefore, the corresponding total cross section
exhibits a clear resonance peak even at the strongest value of the quartic coupling.

The values of the resonance mass we have extracted here are in very
good agreement with earlier determinations \cite{Gerhold:2010bh, Gerhold:2009ub}
where the Higgs boson width has been neglected.
This finding provides confidence to and justifies a posteriori
the Higgs boson mass bounds determined in \cite{Gerhold:2010bh, Gerhold:2009ub}.
In addition,
a comparison of the Higgs boson width with results from perturbation
theory reveals a very good agreement as can be seen in
table~\ref{tab:resonance_mass_compare}.

Figure \ref{fig:chap_conclusion-summary} summarizes the obtained total cross
sections for the three different values of the bare quartic couplings. The displayed
curves correspond to the fits shown in figure
\ref{fig:chap_sm_physics-final_result_cross sections} using the parameterization
of eq.~(\ref{eqn:fitfunction}).
It will be interesting to apply the techniques used in this paper for the
investigation of the Higgs boson mass width in presence of a possible
fourth fermion generation, where non-perturbative effects might
appear for very heavy masses of the fourth generation quarks.
Simulations in this direction have been
started already \cite{Gerhold:2010wv}.
\fig
   {The figure shows the total cross section of the Higgs boson within
   the Higgs-Yukawa model as a function of the momentum $k$.
   With regard to the standard model, this
   cross section is associated to the decay of the Higgs boson into the
   weak gauge bosons $W^{\pm}, Z$.
   The highest peak belongs to the smallest bare quartic coupling
   $\hat \lambda=0.01$ and corresponds to a Higgs boson resonance mass
   of $M_H=248\pm 1$ GeV and a resonance width of $\Gamma_H=6.2\pm 0.9$ GeV.
   The next peak is obtained at $\hat \lambda = 1.0$ and corresponds to
   $M_H=562\pm 2$ GeV and $\Gamma_H=50 \pm 6$ GeV.
   The last peak is associated to infinite bare quartic coupling and
   corresponds to $M_H=618\pm 5$ GeV and $\Gamma_H=60 \pm 6$ GeV.
   In all cases, the resonance width is less than $10 \%$ with respect to
   the resonance mass and thus the corresponding total cross section
   exhibits a clear resonance structure. The energy ranges corresponding
   to the different $\hat \lambda$ values can be
   read off from figure \ref{fig:chap_sm_physics-final_result_cross sections}.}
   {Total cross section of the Higgs boson.}
   {fig:chap_conclusion-summary}
   {0.8\textwidth}
   {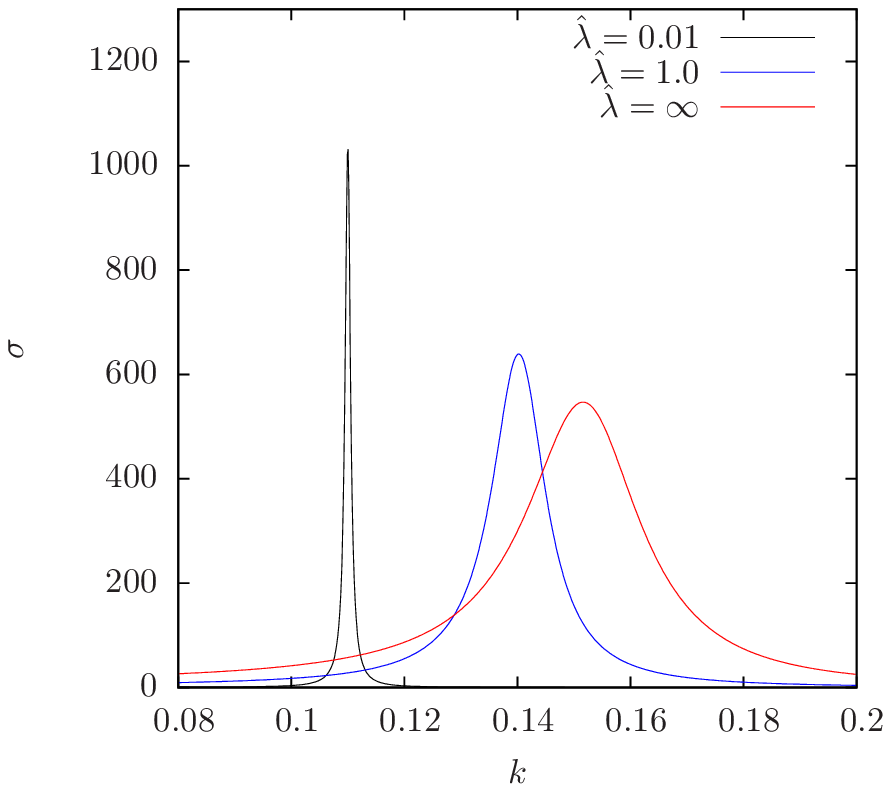}

\section*{Acknowledgments}
We gratefully acknowledge the support of the DFG through the DFG-project {\it Mu932/4-2}.
The numerical computations have been performed on the {\it HP XC4000 System}
at the {Scientific Supercomputing Center Karlsruhe} and on the
{\it SGI system HLRN-II} at the {HLRN Supercomputing Service Berlin-Hannover}.





\clearpage
\bibliographystyle{model1a-num-names}
\bibliography{kallarackal}







\end{document}